%% file: Arxiv2Column.tex
\documentclass[aps,prl,twocolumn,superscriptaddress,groupedaddress,a4paper]{revtex4} 
\usepackage{color}
\usepackage{transparent}
\usepackage{amsmath}
\usepackage{graphicx} 

\newcommand{\Exp}[1]{~\mathrm{exp}\left[#1\right]}
\newcommand{\Noise}{\Lambda^2}
\newcommand{\ExpS}[1]{~e^{#1}}
\newcommand{\Ln}[1]{~\mathrm{ln}\left[#1\right]}
\newcommand{\Cos}[1]{~\mathrm{cos}\left[#1\right]}
\newcommand{\Sin}[1]{~\mathrm{sin}\left[#1\right]}
\newcommand{\E}[1]{\left<#1\right>}

\newcommand{\PartialD}[2]{\frac{\partial#1}{\partial#2}}

\newcommand{\MSpartialD}[3]{\frac{\partial^2#1}{\partial#2\partial#3}}

\newcommand{\Eqn}[1]{Eq.~(\ref{#1})}
\newcommand{\Fig}[1]{Fig.~\ref{#1}}
\newcommand{\Figu}[2]{Fig.~\ref{#1}.#2}

\newcommand{\enquote}[1]{``#1''}
\newcommand{\RegBD}{\mathcal{B}\oplus\mathcal{D}}

\begin{document}

\title{Theoretical optimal modulation frequencies for scattering parameter estimation and ballistic photon filtering in diffusive media}

\author{Swapnesh Panigrahi}

\affiliation{Institut de Physique de Rennes, CNRS, Université de Rennes 1, Campus de Beaulieu, 35 042 Rennes, France}

\author{Julien Fade}
\affiliation{Institut de Physique de Rennes, CNRS, Université de Rennes 1, Campus de Beaulieu, 35 042 Rennes, France}
\author{Hema Ramachandran}
\affiliation{Raman Research Institute, CV Raman Avenue, Sadashivanagar, 560 080 Bangalore, India}
\author{Mehdi Alouini}
\affiliation{Institut de Physique de Rennes, CNRS, Université de Rennes 1, Campus de Beaulieu, 35 042 Rennes, France}



\begin{abstract}
The efficiency of using intensity modulated light for estimation of scattering properties of a turbid medium and for ballistic photon discrimination is theoretically quantified in this article. Using the diffusion model for modulated photon transport and considering a noisy quadrature demodulation scheme, the minimum-variance bounds on estimation of parameters of interest are analytically derived and analyzed. The existence of a variance-minimizing optimal modulation frequency is shown and its evolution with the properties of the intervening medium is derived and studied. Furthermore, a metric is defined to quantify the efficiency of ballistic photon filtering which may be sought when imaging through turbid media. The analytical derivation of this metric shows that the minimum modulation frequency required to attain significant ballistic discrimination depends only on the reduced scattering coefficient of the medium in a linear fashion for a highly scattering medium.
\end{abstract}

\maketitle

\section{Introduction}

Imaging through and within turbid media is an area of interest that has tremendous application in medical diagnostics \cite{Boas2001}, underwater vision \cite{Schechner2004}, imaging through colloids \cite{Ramachandran1998} and transportation and navigational aids \cite{Watkins2000,Hautiere2007}. Light traveling through a complex medium with randomly distributed positions and refractive indices undergoes absorption and random scattering and loses the spatial and temporal information of its source. The photons that undergo such multiple scattering are labeled as diffusive photons. A small fraction of the total photons called ballistic photons, undergo forward scattering to reach a detector and they retain the information of its source. It is of wide interest to discriminate the ballistic photons from the diffused photons for resolution enhanced imaging through turbid media. However, the diffuse light that strongly depends on the properties of the scattering medium can be used to deduce various parameters related to the medium itself. Thus, imaging in turbid media can be classified into two broad categories: parameter estimation using only diffused light to obtain image of heterogeneities in the turbid media and filtering of ballistic photons from diffused light for high resolution imaging through turbid media.

\paragraph{Parameter estimation}
In parameter estimation, the physical properties of the intervening media are of interest and controlled light sources may be used to estimate the scattering and absorption parameters \cite{Arridge1999,Brewster1995}. Diffused optical imaging which has been widely studied and applied in medical imaging \cite{Tromberg2000} is used to estimate the scattering and absorption coefficient in the intervening media for detection of malignant tissues in breast \cite{Colak1999} or for brain imaging \cite{Strangman2002}. In such cases the time dependent solution to diffusion theory is used to model the transport of pulsed light \cite{Benaron1993,Cubeddu1996} or modulated light \cite{Gratton1997,Tromberg1993,Jiang1995,Pogue1994} through the scattering media. The precision of estimation of the parameters is crucial in this case. In practice, intensity modulated light with diffusion theory are widely used in temporal frequency-domain photon migration, where an intensity modulated light with modulation frequencies of a few hundred MHz is transmitted through a diffusive medium. According to the time dependent solution of the diffusion theory, at these frequencies, the wave number of the density wave traveling through the medium is dependent on the scattering and absorption properties of the medium. As a result, the detected modulated light has reduced modulation index and an additional phase, both of which depend on the scattering properties of the medium and its thickness. As the diffusion theory provides an analytical model for the change of modulation index and phase of the modulated light, the parameters of the intervening medium could be estimated by using a single or multiple modulation frequencies.

\paragraph{Ballistic discrimination}
On the other hand, ballistic filtering is used when the spatial resolution of objects embedded in scattering media is of interest\cite{Farsiu2007,Wang1991}. For instance, for industrial applications where it is required to image objects embedded in colloidal system, or in navigation where vision through fog can be efficiently achieved with ballistic filtering. The problem of imaging through such media has been addressed using various techniques that essentially rely on discriminating forward scattered ballistic photons from multiply scattered (diffuse) photons traversing through a turbid medium. For example, time gated imaging \cite{Sedarsky2011,Berg1993}, polarization gated imaging \cite{Emile1996}, intensity modulation imaging \cite{Mullen2004} etc., where information carried by the ballistic photons is filtered from the diffuse light. It is challenging to efficiently detect the ballistic photons that have low signal to noise ratio as compared to the diffused photons, especially in highly scattering media. Ballistic photon filtering can also be achieved by an intensity modulation imaging scheme by discriminating them from the diffused light that arrives with reduced modulation index and additional phase.\\ 


In this article, we analyze the widely used diffusion approximation for transport of modulated light, from an information theoretical point of view for efficient parameter estimation and for ballistic photons discrimination. Assuming a general scheme of optical quadrature demodulation detection and a corresponding noise model, this approach provides a rigorous insight into the maximal precision of estimation of parameters using the diffusion equation, thereby providing a robust theoretical argument for choosing the frequency of modulation suited to the experiment at hand. Similarly, for ballistic filtering, we will present an information theoretic performance metric and study the optimum ballistic discrimination efficiency that can be achieved under this imaging scheme.

In section \ref{sec:DiffusionModel}, we provide a brief description of the use of diffusion equation, the transport of modulated light in diffused media and the imaging scheme along with the parameters that will be used in the theoretical analysis. In section \ref{sec:noiseModel}, a generalized demodulation scheme is defined and a noise model is presented. Then, in section \ref{sec:CRB}, the noise model and the diffusion theory are used to calculate the Cramer-Rao bound for parameter estimation. Finally, in section \ref{sec:ballisticFiltering}, the ballistic filtering efficiency is analyzed and discussed with respect to real field situations.

\section{Imaging scheme and diffusion model} \label{sec:DiffusionModel}

    \begin{figure}[htbp]
    \centering
    \def\svgwidth{0.9\columnwidth}
    \fbox{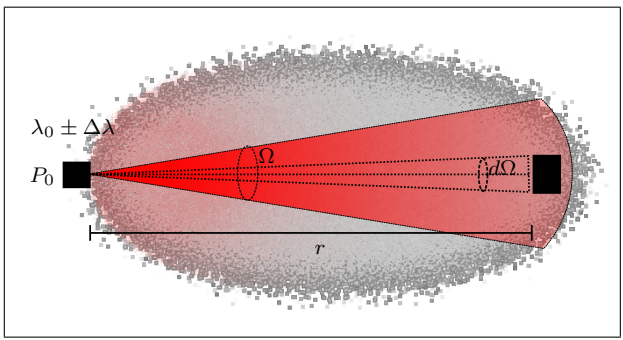}
    \caption{Imaging scheme: A directional source of light with power, $P_0$, forward cone solid angle, $\Omega$, having a limited spectral width ($\lambda \pm \Delta\lambda$) is detected at a distance $r$ by a detector that subtends an angle $d\Omega$ from the source.}
    \label{fig:schema}
    \end{figure}


    The diffusion theory for photon transport provides a simple, fast and analytical method for modeling the light propagation through various turbid media. The properties of intensity modulated light through a diffusive medium have been well studied and reviewed 
    \cite{Patterson1989,Martelli1997a,Fishkin1993,Gratton1997}. Here, we present a brief introduction underlining concepts that are relevant in the context of this article. We will follow the modeling of the intensity modulated light in diffusive media as derived in \cite{Tromberg1993} and many others \cite{Fishkin1993,Jacques2008} in the context of diffuse optical imaging. We will use the formulations previously derived by the cited authors to parameterize our detection scheme and then use information theoretical tools to derive simple rules for choosing frequency operating points to efficiently use intensity modulation light depending on the properties of the medium.

\subsection{Diffusion model}

The diffusion model arises when the photons are allowed to perform a random walk, diffusing from high photon density regions to low photon density regions. The theory has proved efficient when modeling light in a predominantly scattering medium and when detection is carried out sufficiently away from a point source. The efficiency of the diffusion theory has been studied alongside Monte-Carlo simulation and shown to have well acceptable accuracy in the domain of its validity \cite{Flock1989}. Let us now identify some parameters that are important for describing transport in a scattering medium. The scattering length ($l_\sigma$) is defined as the mean distance traveled by a photon before the scatter. Its inverse, the reduced scattering coefficient is denoted by $\sigma$ and defined as $\sigma = \mu_s (1-g) $, where $\mu_s$ is the scattering coefficient and $g$, the anisotropy factor, is the mean cosine of the scattering phase function \cite{wang2007biomedical,ishimaru1999wave}.

Another important length scale is the transport mean free path ($MFP$) denoted by $l_{*}$, which is the mean distance traveled by photons before they lose their initial directional information. The diffusion theory also includes two other constants: the diffusion length ($D$) defined as $D=l_{*}/3$ and the optical penetration depth ($\delta$), which is the inverse of the effective attenuation coefficient ($\sqrt{\mu/D}$) in diffused medium. Finally, the absorption coefficient ($\mu$) is associated with an absorption length. For ease of reading, the corresponding definitions are tabulated in Tab~\ref{tab:lengthSclaes}.

In addition, for the analysis in the following sections we will use various dimensionless constants that are also given in the right side of Tab~\ref{tab:lengthSclaes}. The dimensionless parameters $R_\delta = r/\delta$ and $R_{*} =r/l_* $ correspond to the effective optical attenuation of diffused light and effective optical attenuation of ballistic light, respectively. The parameter $q$ is related to the angular frequency of modulation and the non-trivial form of the reparameterization will be justified in following sections.

\subsection{Imaging scheme}
\begin{table*}
    \centering
    \caption{Symbols, definitions and dimensionless parameters}\label{tab:lengthSclaes}
    \begin{tabular}{ l l | l l }        
    \hline
    Meaning & Symbol & Param. & Defn.	\\
    \hline
    Transport MFP & $l_{*} = 1/(\mu + \sigma)$ & $R_{*}$ & $r/l_{*}$ \\ 
    Diffusion length & $D = l_{*}/3$ & -- & -- \\
    Optical penetration depth & $\delta =[D/\mu]^{\frac{1}{2}} =[3\mu(\mu + \sigma)]^{-\frac{1}{2}}$  & $R_\delta$ & $r/\delta$ \\
    Distance of propagation & $r$ & --&-- \\
    Angular modulation frequency & $\omega$ & $q$ & $\sqrt{\frac{1+\sqrt{1+(\omega/\mu c)^2}}{2}}$  \\    
    Effective refractive index & $n$ & & \\
    Speed of light in medium & $c$ & -- & -- \\
    \hline 
    \end{tabular}    
\end{table*}

The imaging scheme considered in this article includes a directional point source of light with a forward cone solid angle $\Omega$ and that subtends a solid angle of $d\Omega$ at the detector which is placed at a distance $r$ from the source. For the sake of simplicity, we shall consider a source of limited spectral range $\lambda_0 \pm \Delta\lambda$, so that the above diffusion parameters can be considered as constant over the considered spectral range. The schematic in \Fig{fig:schema} illustrates the scenario. In the presence of an intervening scattering medium, the net intensity of ballistic photons reaching the detector of collection area $d\Omega r^2$ is proportional to the total power $P_0$ emitted by the source, and is given by the Beer's law as $I_B = \xi P_{0} \ExpS{-(\mu + \sigma)r} \frac{d\Omega r^2}{\Omega r^2} = \xi P_{0} \ExpS{-(\mu + \sigma)r} \frac{d\Omega}{\Omega}$. The scaling factor $\xi$ represents the overall detector efficiency on the considered spectral range. Similarly, according to the steady state solution of the diffusion theory, the diffuse photon intensity reaching the detector is also proportional to the total power emitted by source such that $I_D = \xi P_{0} \ExpS{-r/\delta} \frac{d\Omega r^2}{4\pi D r} = \xi P_{0} \ExpS{-r/\delta} \frac{d\Omega}{4\pi} \frac{r}{D}$ \cite{Tromberg1993,Jacques2008}. The expressions of intensities of ballistic and diffused light show that there are clearly two important length scales to be considered, namely, the transport MFP ($l_{*}$) and the optical penetration depth ($\delta$). Considering only the above two classes of photons, it is possible to obtain an order of magnitude value of the ratio ($\alpha$) of ballistic photons to diffuse photons reaching the detector as 

    \begin{equation}
    \alpha = \frac{I_B}{I_D} =  \Omega' \frac{D}{r} \ExpS{-r(\mu + \sigma - 1/\delta) } = \Omega' \frac{e^{R_\delta-R_{*}}}{3 R_{*}}, 
    \end{equation}
where we use dimensionless variables $R_{*}$, $R_\delta$ and $\Omega'= 4\pi/\Omega$.

\subsection{Modulated light in diffused media}

The propagation of sinusoidally modulated light through a scattering medium has been modeled using diffusion theory and it has been shown that the transport of modulated light behaves as photon density waves whose properties are dependent on the properties of the medium \cite{Tromberg1993,Fishkin1993,Martelli1997a,Gratton1997}. In this article, we consider an intensity modulated source of light having modulation angular frequency $\omega$, modulation index $M$ and instantaneous intensity $i(t) = I_0 ( 1+ M \Cos{\omega t}) $. The ballistic light that follows Beer-Lambert's law, is only attenuated and reaches the detector with instantaneous intensity $i_b(t) = I_B (1+ m_B \Cos{\omega t})$ without any change in received modulation index, $m_B = M$.

However, the time dependent solution of the photon diffusion theory shows that the modulated light traversing through a scattering medium is received at the detector with reduced modulation index and additional phase \cite{Tromberg1993}. Then, the instantaneous diffuse light intensity received at the detector is $i_d(t) = I_D ( 1+ m_D \Cos{\omega t + \Delta\phi}) $. Without derivation, we present the expression of the reduced modulation index $m_D$ and the phase $\Delta\phi$ which is identically reported in \cite{Tromberg1993,Jacques2008}:

    \begin{widetext}
    \begin{subequations}\label{eq:dispersion}
    \begin{eqnarray}     
    m_D &= M \Exp{- r [3\mu(\mu+\sigma)]^{\frac{1}{2}}  \left( \sqrt{\frac{1+\sqrt{1+ \left(\omega/\mu c\right)^2}}{2}} -1 \right)} = M \ExpS{-R_\delta  (q-1)},   \\    
    \Delta\phi &= r \sqrt{3\mu(\mu+\sigma)}  \sqrt{\frac{-1+\sqrt{1+ \left(\omega/\mu c\right)^2}}{2}} = R_\delta \sqrt{q^2-1}.    
    \end{eqnarray}
    \end{subequations}
    \end{widetext}
    where the parameter $q = \sqrt{(1+\sqrt{1+(\omega/\mu c)^2})/2}$,
    is related to the angular frequency of modulation and ranges
    between $[1,\infty)$ when $\omega\in [0,\infty)$. Although the
    physical interpretation of this parameter is not straightforward,
    we will see later that $q R_\delta$ can be identified as a
    dimensionless, frequency-dependent, effective attenuation of the
    diffused light. In the remainder of this article, we define
    $\beta=m_B/m_D$ as the ratio of the modulation indices of
    ballistic light to diffuse light. The expressions of the
    intensity, phase and modulation index of ballistic and diffuse
    light components are recalled in Table \ref{tab:BvsD}.

 \begin{table*}
    \centering
    \caption{Expressions of intensity, modulation index and relative phase for the ballistic and diffuse components of the detected light. \label{tab:BvsD}}
    \begin{tabular}{ l  c  l  r}        
       \hline
       Detection & Ballistic photons & Diffuse photons & Ratio (Ballistic/Diffuse)\\
       \hline
       Relative phase ($\Delta\phi$) & 0 & $R_\delta \sqrt{2(q^2-1)}$ & --\\
       Modulation index & $m_B = M$ & $m_D = M e^{-R_\delta \left(q-1\right) }$ & $\beta =  e^{R_\delta \left(q-1\right) }$\\
       Intensity  & $I_B = \xi \frac{d\Omega}{\Omega} P_{0} \ExpS{-R_{*}}$ & $I_D = \xi \frac{d\Omega}{4\pi} 3 P_0 R_{*} \ExpS{-R_\delta}$ & $\alpha = \Omega' \frac{e^{R_\delta-R_{*}}}{3 R_{*}} $\\
       \hline   
    \end{tabular}
    \end{table*}

It is quite straightforward to see that using the above equations, the parameters $R_\delta$ and $q$ can be estimated when the modulation index and phase of the diffuse light are accurately detected. The model can indeed be inverted as shown in \Eqn{eq:InvEsti}: 
\begin{subequations}\label{eq:InvEsti}
    \begin{eqnarray} 
    R_\delta = \frac{\Delta\phi^2-2\Ln{\beta}^2}{4 \Ln{\beta}},\\
    q = \frac{\Delta\phi^2 + 2 \Ln{\beta}^2}{\Delta\phi^2 - 2 \Ln{\beta}^2}.
    \end{eqnarray}
    \end{subequations}

Thus, the above formulation provides a simple analytical method for estimation of the scattering and absorption parameters of the scattering medium using only diffuse light and a modulated light source. In practice, especially in diffuse optical imaging, the modulation frequency is scanned to obtain corresponding values of modulation index and phase. Then, a non-linear fit of the theoretical prediction with the data provides the estimates for the scattering and absorption properties of the medium \cite{Tromberg1993}. The effect of scattering media on modulation index and phase can also be exploited to attain discrimination of ballistic photons that retain the modulation index and phase. These application scenarios can be analyzed from an information theoretical point of view for a well-defined detection technique. The detection of the modulation index and the phase can be performed in various ways. Generally in a demodulation scheme, the amplitude, phase and mean intensity are recorded and then the modulation index can be easily estimated. Quadrature demodulation is one of the simplest and the most widely used scheme for demodulation. It avoids, in particular, phase tracking of the incoming signal which brings additional noise contributions. In the following section we look at a quadrature detection scheme and derive the noise model for the detection.

\section{Quadrature detection scheme and noise model} \label{sec:noiseModel}

    \begin{figure}
        \centering        
        \def\svgwidth{0.5\columnwidth}
        \fbox{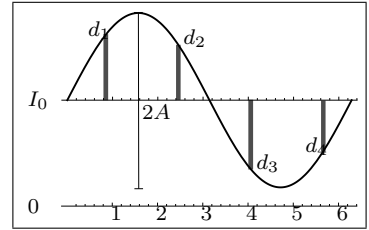}
        \caption{Illustration of the optical signal received at the detector. As an example, four samples are shown here to form the quadrature components which, consequently, can be used for estimation of amplitude and phase of the signal.} \label{fig:Sin}
    \end{figure}

Demodulation of a sinusoidally modulated signal can be performed by product detection, where the signal is continuously multiplied with a sine and cosine of the same frequency to obtain the quadrature amplitudes. This quadrature detection can be interpreted as follows. Let us consider a detection system where a sinusoidal signal of angular frequency $\omega$, modulation index $M$ and mean intensity $I_0$ is sampled $N$ times in a period producing photon count or grey level data $d_i$ ($i=0$ to $N$). Then the quadrature components can be estimated by applying sinusoidal weighting of the data points as $U = \sum_{i=1}^N d_i \Cos{\omega t_i + \phi} = A \Cos{\phi}$ and $V = \sum_{i=1}^N d_i \Sin{\omega t_i + \phi}= A \Sin{\phi}$, for $N\ge2$. From the obtained variables $U$ and $V$, the amplitude $A = MI_0/2$ and phase $\phi$ of the demodulated signal are respectively given by $A = \sqrt{U^2 +V^2}$ and $\phi = \text{tan}^{-1}[V/U]$.

The intensity statistics of the random variables $U$ and $V$ can be derived by knowing that the photon count within an infinitesimal sampling window at time $t_i$ may follow Poisson distribution with mean and variance equal to the intensity received at the sampling window, which is $d_i$ in the current notation. Then, a weighted addition of the Poisson distributed data is also Poissonian, indicating that $U$ and $V$ will also be distributed as Poisson random variables with variance $I_0/2$ (see Appendix~\ref{ax:Noise_variance}). For high intensities, the Poisson distribution can be well approximated by a Gaussian distribution ($\mathcal{N}$) with variance equal to the mean. Thus, we approximate that $U$ and $V$ are distributed as $\mathcal{N}(A \Cos{\phi},\Noise)$ and $\mathcal{N}(A \Sin{\phi},\Noise)$, respectively, with $\Noise = I_0/2$ and $\mathcal{N}(\bar{x},var(x))$ denoting the normal distribution with mean $\bar{x}$ and variance $var(x)$. Knowing the distribution of the quadrature components, it then possible to derive the joint probability density function (PDF) of the random variables, $Z = \sqrt{U^2 + V^2}$ and $\Psi = \text{tan}^{-1}(V/U)$, by applying the appropriate change of variables to the Gaussian distribution, as given in the Appendix~\ref{ax:JointPDF}. The expected values of the random variables $Z$, $\Psi$ and their variance form the parameter vector $\boldsymbol{\theta}' = [A,\phi,\Noise]$, where $2A = M I_0$ is the peak-to-peak amplitude as shown in the schematic of \Fig{fig:Sin}. Then, the joint PDF of the amplitude and phase is
    \begin{equation} \label{eq:pdf}
    \begin{aligned}
    &P_{Z,\Psi}(z,\psi| A, \phi,\Noise) = \\ & \frac{z}{2 \pi \Noise} \Exp{-\frac{1}{2\Noise} (z^2 + A^2 - 2 z A \Cos{\psi - \phi})}.
    \end{aligned}
    \end{equation}

Quadrature demodulation cameras are in widespread use, especially in 3D imaging. For instance, novel demodulation cameras like time-of-flight (TOF) cameras collect four samples per period to obtain the quadratures $U = d_1-d_3$ and $V = d_2-d_4$ at frequencies up to 20 MHz \cite{Lange2001}. Information theoretical studies have been made to accurately determine the phase which contains the depth information of a scene using the above PDF \cite{Mufti2009}. In the following, we will use Fisher information (FI) and Cramer-Rao bound (CRB) to analyze how well the diffusion parameters of the scattering medium can be estimated using modulated light and the quadrature detection technique.

\subsection{Fisher information}
Fisher information (FI) is a useful and efficient method of quantifying the precision with which the unknown parameters in a data model can be estimated. It can be further used to analyze the behavior of the covariance of estimators and its dependency on other parameters. FI is ultimately related to the minimum covariance bound on the unbiased estimation of the unknown parameters. Here, we will use the FI to quantitatively derive simple rules that must be taken into account when using intensity modulated light in diffusion approximation for estimation of scattering properties of a medium and for ballistic filtering applications.

The expected Fisher information matrix (FIM) for a parameter vector ($\boldsymbol{\eta}$), is defined as $ \mathcal{F}(\boldsymbol{\eta})_{ij} = - \E{\MSpartialD{\Ln{P(x|\boldsymbol{\eta})}}{\eta_i}{\eta_j} }$, where $\E{.}$ denotes the expectation value. We present below the FIM for the detection procedure described above with respect to the parameter vector $\boldsymbol{\theta}' = [A,\phi,\Noise]$. It is straightforward to show that the FI reads in that case

\begin{equation}
    \mathcal{F}(\boldsymbol{\theta}') = 
    \left[
    \begin{array}{ccc}
    \frac{1}{\Lambda^2 } & 0 & 0 \\
    0 & \frac{A^2}{\Lambda^2 } & 0 \\
    0 & 0 & \frac{1}{\Lambda ^4} \\
    \end{array}
    \right].
\end{equation}

    The above FIM is diagonal, which shows that the three parameters in $\boldsymbol{\theta'}$ can be estimated independently. Moreover, the precision in estimation of the phase increases with the amplitude of the signal. 
    The detection technique described in the preceding section is not just limited to detector arrays, but is generally adopted in lock-in detection. In this detection scheme, the recorded amplitude and phase of the diffuse light through a scattering media can be modeled by the diffusion theory as presented in preceding sections. Using this model, the noise model of the detection can be reparametrized as shown below and consequently, estimation precision of the parameters can be calculated.

\subsection{Reparametrization of noise model using diffusion theory}
Let us consider the effect of diffused light only, and obtain the FIM with respect to the new set of parameters $\boldsymbol{\theta}= [M,R_\delta,R_{*}]$ to deduce some insight into the precision of estimation of each respective parameter and its dependency on other parameters. The FI with respect to the new parameters can be obtained by calculating the Jacobian matrix, $J_{\cal D}$, of the transformation $\boldsymbol{\theta'} \to \boldsymbol{\theta}$ in the presence of diffuse light only, as denoted by the suffix ${\cal D}$. The $i^{th},j^{th}$ component of this matrix is $\bigl[J_{\cal D}\bigr]_{i,j}=\PartialD{\theta'_i}{\theta_j}$. Given the modulation index, phase and the intensity expected for diffuse light from the diffusion model, the amplitude, phase and variance can be written as $A_D = m_D I_D/2$, $\Delta\phi = R_\delta \sqrt{2(q^2-1)}$ and $\Noise_D = I_D/2 = 3 S_0 R_{*} \ExpS{-R_\delta}/2$, where we have set $S_0= \xi d\Omega P_0 /4\pi$. The Jacobian is then calculated to obtain the FIM with respect to parameter vector $\boldsymbol{\theta}$ using the transformation $\mathcal{F}_{\cal D}(\boldsymbol{\theta}) = J_{\cal D}^T \mathcal{F}(\boldsymbol{\theta'}) J_{\cal D}$. Both, the Jacobian and the FIM $\mathcal{F}_{\cal D}(\boldsymbol{\theta})$ are shown in Appendices~\ref{ax:Jacobian} and~\ref{ax:FIM}, for reference. It is worth mentioning here that in applications where the SNR of the amplitude and phase are important, the noise model presented here can be used to obtain the SNR of detected amplitude and phase as a function of frequency by using the above relations. However, we will focus on deriving and analyzing the minimum variance bounds on estimation of scattering parameters as presented in the following section.

\section{Maximal precision in estimation of scattering parameters in diffuse optical imaging} \label{sec:CRB}

\subsection{Lower bound on estimation variance}

The above re-parameterization allows us to study the variance bound in estimation of $\boldsymbol{\theta}$ with respect to the frequency of modulation represented by the variable $q$. According to the Cramer-Rao theorem, for any unbiased estimator $\hat{\boldsymbol{\theta}}$ of the parameter vector $\boldsymbol{\theta}$, one has $\E{w}\E{\hat{\boldsymbol{\theta}} \hat{\boldsymbol{\theta}}^T} \E{w}^T \geq \E{w} \mathcal{F}(\boldsymbol{\theta})^{-1} \E{w}^T$. Thus, the Cramer-Rao bound (CRB) provides a minimum covariance bound that can be reached by an efficient estimation technique. Generically, it is not guaranteed that such a efficient estimator always exists, however, Maximum Likelihood (ML) estimators have been shown to be asymptotically unbiased and efficient for large collection of data, especially for exponential family of distributions \cite{garthwaite2002statistical}. 
The exact forms of ML estimators in this case are solutions to transcendental equations and remain out of the scope of this article. Instead, let us analyze the lower covariance bound to reveal some simple conclusions and rules for the estimation problem. We provide the bound as $CRB_{\cal D}(\boldsymbol{\theta}) = \mathcal{F}_{\cal D}(\boldsymbol{\theta})^{-1}$ for the invertible matrix $\mathcal{F}_{\cal D}(\boldsymbol{\theta})$ which is shown in the Appendix~\ref{ax:FIM}:
\begin{widetext}
\begin{equation} \label{eq:CRB}    
    CRB_{\cal D} (\boldsymbol{\theta}) = 
    \left[
    \begin{array}{ccc}
     M^2+\frac{4 e^{(-1+2 q) R_{\delta }} q}{3 (1+q) R_{\text{*}} S_0} & \frac{2 e^{(-1+2 q) R_{\delta }}}{3 M (1+q) R_{\text{*}} S_0} & -M R_{\text{*}}+\frac{2 e^{(-1+2 q) R_{\delta }}}{3 M (1+q) S_0} \\
     \frac{2 e^{(-1+2 q) R_{\delta }}}{3 M (1+q) R_{\text{*}} S_0} & \frac{2 e^{(-1+2 q) R_{\delta }}}{3 M^2 \left(-1+q^2\right) R_{\text{*}} S_0} & \frac{2 e^{(-1+2 q) R_{\delta }}}{3 M^2 \left(-1+q^2\right) S_0} \\
     -M R_{\text{*}}+\frac{2 e^{(-1+2 q) R_{\delta }}}{3 M (1+q) S_0} & \frac{2 e^{(-1+2 q) R_{\delta }}}{3 M^2 \left(-1+q^2\right) S_0} & R_{\text{*}}^2+\frac{2 e^{(-1+2 q) R_{\delta }} R_{\text{*}}}{3 M^2 \left(-1+q^2\right) S_0} \\
    \end{array}
    \right]    
    \end{equation}
\end{widetext}    

\begin{figure*}
        \centering        
        \def\svgwidth{0.8\linewidth}
        \fbox{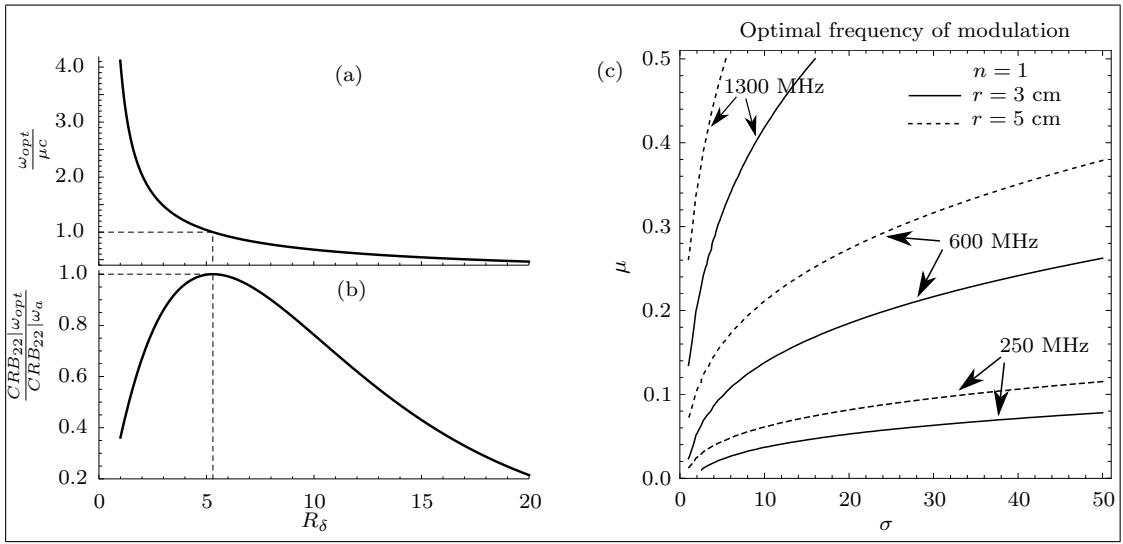}
        \caption{(a) Ratio of the optimal frequency $\omega_{opt}$ to
          standard operating frequency $\omega_a=\mu c$ as a function
          of the normalized optical attenuation $R_\delta$. (b) Loss
          in precision in the estimation of $R_\delta$ using
          modulation angular frequency $\omega_a$ as opposed to
          $\omega_{opt}$ as a function of $R_\delta$. (c) Contours of
          optimal frequency of modulation as a function of absorption
          coefficient $\mu$ and reduced scattering coefficient
          $\sigma$ and the detection distance $r$ for an intervening
          medium having refractive index 1. The frequencies can be
          scaled down by a factor of $n$ for a medium with refractive
          index $n$. } \label{fig:q_opt}
    \end{figure*}

It is clearly seen that the variances are functions of the frequency of modulation (represented by $q$). Let us first look at the variance in estimation for the parameters $R_\delta$ and $R_{*}$. The variance of $R_{*}$ increases with its mean value and has an additional frequency dependent term, while the variance of $R_\delta$ decreases with the $R_{*}$. The dependence of both the parameters on frequency is the same and has a functional form $\ExpS{(-1+2q)R_\delta}/(-1+q^2)$. A simple calculus shows that this function reaches a minimum at $q_{opt}=\bigl(1+\sqrt{1+4 R_\delta^2}\bigr)/2 R_\delta$. This indicates that there exists an optimal angular frequency $\omega_{opt}$ around which the variance of the estimation is minimum. This optimal angular frequency depends only on $R_\delta$ and is given by the following expression
    \begin{equation} \label{eq:OptimalFreq}
      \frac{\omega_{opt}}{\mu c} = \frac{ \sqrt{2}\left[ 1+\bigl[1+4 R_{\delta }^2\bigr]^{\frac{1}{2}}+R_{\delta }^2 \bigl( 3+\bigl[1+4 R_{\delta }^2\bigr]^{\frac{1}{2}}\bigr) \right]^{\frac{1}{2}} }{R_{\delta }^2},
    \end{equation} 
    whose evolution is plotted in \Figu{fig:q_opt}{a}. In the next
    subsection we analyze the properties and evolution of the optimal
    frequency $\omega_{opt}$ obtained for the estimation of scattering
    parameters.

\subsection{Optimal operating frequency}

The expression of the variance-minimizing optimal angular frequency $\omega_{opt}$ has a non-trivial form that depends on the normalized optical attenuation $R_\delta=r/\delta$, which basically represents the inverse of an overall visibility factor for the diffuse photons. The existence of an optimal operating point has interesting consequences, especially in the context of diffused optical imaging, where the SNR of the images is an important limiting factor. Generally, in such applications, the operating frequency, denoted here by $\omega_a$, is chosen such that $\omega_a/\mu c = 1$ \cite{Jacques2008}. This is indeed a reasonable choice of operating frequency, justified by the fact that at smaller frequencies on the one hand, the phase change due to diffusion is too small to be detected. On the other hand at much higher frequencies, the change in phase becomes comparatively insensitive to further increase in frequency, thereby not bringing any further improvement in the estimation of the medium parameters \cite{Jacques2008}.

To understand the loss incurred by an imperfect choice of operating frequency, we compare the performance of estimation provided by the two frequencies, $\omega_{opt}$ and $\omega_a$. We plot the ratio of the variance in the estimation of $R_\delta$ obtained when using $\omega_{opt}$ against $\omega_{a}$ in \Figu{fig:q_opt}{b}. The figure provides a quantitative study of the loss in optimal precision in using $\omega_a$ as opposed to using $\omega_{opt}$. The precision of estimation are equal only when $R_\delta = 5.3$, at which point $\omega_{opt}=\omega_{a}$, as indicated by the dashed lines in the figures. For example, taking typical values of scattering parameters valid in tissues, such that $\mu = 0.1$ cm$^{-1}$ and $\sigma = 10$~cm$^{-1}$, the two operating points are the same only for a detector placed as distance of $3$ cm. The standard operating frequency $\omega_a$ is independent of the propagation distance $r$ and of $\sigma$, depending only on the absorption coefficient $\mu$. As illustrated in \Figu{fig:q_opt}{a}, for a detection carried out at a distance higher than $3$ cm, the optimal frequency is in fact smaller which may reduce the cost and complexity of the electronics required at high frequencies while providing better results. On the other hand, if the detection is carried out closer than $3$ cm, higher operating frequencies are suggested as compared to standard operational frequency. 

Similarly, since $R_\delta$ is a function of $\sigma$, $\mu$ and $r$, the dependency of the optimal modulation frequency on these three parameters can be analyzed from the contour plot shown in \Figu{fig:q_opt}{c}. Firstly, for fixed values of $\sigma$ and $\mu$, the optimal operating frequency decreases with increase in $r$. Secondly, for fixed values of $r$ and $\mu$, the optimal frequency is seen to decrease with the increase in $\sigma$. The above two dependencies can be interpreted by noticing that longer distance of travel or higher scattering coefficient would allow for a larger amount of multiple scattering events leading to greater modification of the phase and the modulation index of the diffused light, for a given coefficient $\mu$.

Finally, the optimal frequency is seen to increase with the increase in $\mu$, for fixed values of $r$ and $\sigma$. An increase in absorption would lead to smaller amplitude of modulation at the detector. To compensate for this loss that occurs at a rate of $\mu c$, the photon density arrival rate should be increased leading to a increase in optimal frequency of operation. It can also be noticed that the change of optimal operating frequency is more sensitive to a change in absorption coefficient than to a change in $\sigma$ or $r$. The above conclusions can also be easily obtained by noticing that equation \Eqn{eq:OptimalFreq} can be approximated to $\omega_{opt} = 2 \mu c /\sqrt{R_\delta}= 2 \sqrt{\mu} c  [r\sqrt{3(1+\sigma/\mu)}]^{-1/2}$, for sufficiently large value of $R_\delta$. It is interesting to notice at this level that the above analytical deductions about the existence of an optimal operating frequency and its evolution with scattering parameters are qualitatively in very good agreement with previous numerical simulations and experimental studies presented in \cite{kimoptimal2008,gufrequency-domain2007,toronovoptimization2003}.

More generally, the above results show that our analysis, based on a diffusion model for the propagation of photons coupled with an information theoretical approach, makes it possible to account for several competing phenomena involved in light diffusion (e.g., the variance of detection decreases with $R_\delta$ but is independent of $\omega$, whereas the modulation index and the phase depend on $R_\delta$ but with rates of change that are functions of the angular frequency). This analysis is able to provide the functional dependence of the optimal operating frequency with respect to properties of the medium under consideration and practical indications concerning the optimal experimental parameters for the estimation task considered. The conclusion obtained appears to be more specific than the usual rule of thumb used in similar experiments, as it reveals the optimal operating frequency and provides insight into the various competing phenomena that produce it.

It is possible to extend such an information theoretical analysis towards another problem of ballistic photon filtering in diffusive media, where the goal is to efficiently discriminate the ballistic photons from the diffuse photons. In the next section we will introduce a performance metric for ballistic discrimination and present the optimal operating point based on the information theoretical approach coherent with the previous discussions.

\section{Ballistic filtering} \label{sec:ballisticFiltering}

The importance of ballistic discrimination has been discussed in the
introduction section, for instance when imaging objects embedded in
nebulous media like fog. As noted above, when high-frequency modulated
light propagates in a turbid medium, the ballistic photons and
diffusive photons have different transport properties in terms of
modulation index and phase, which can be exploited by a demodulating
detector to attain discrimination of ballistic photons.

By contrast, such a physical ballistic filtering effect cannot be
obtained with a standard intensity camera, or with low-frequency
modulation/lock-in detection. In both cases, detection of an extremely
small ballistic contribution over a spatially uniform diffuse
illumination would be possible only at the expense of a dramatic
increase in the detector dynamics or acquisition time.

In this section, we investigate the conditions required to obtain
significant ballistic filtering with modulated light in a turbid
medium. More particularly, we derive the minimum modulation frequency
required to attain ballistic filtering irrespective of photon budget
and exposure time.

\subsection{Gain definition for ballistic filtering efficiency}

We will again resort to FI to define a ballistic filtering efficiency
as the gain in information provided by the ballistic light for the
estimation of the modulation index $M$ of the light source. Let us
consider a quadrature demodulation camera that receives ballistic
light over the diffused light on a set of pixels denoted $\RegBD$ and
another set of pixels that receive only diffused light at a region
denoted by $\mathcal{D}$. In most cases, the contrast between these
two regions is marginal because fewer ballistic photons reach the
detector. To quantify this contrast in a demodulation scheme, we
define the ballistic discrimination efficiency or gain as the ratio of
FIs in the estimation of the actual modulation index $M$ of the source
when using data from region $\RegBD$ as opposed to $\mathcal{D}$. The
gain in information, denoted by $\mathcal{G}_{bf}$ is thus defined as
\begin{equation}\label{def_gain_bf} \mathcal{G}_{bf}=\frac{[\mathcal{F}_{\RegBD}(\boldsymbol{\theta})]_{11}}{[\mathcal{F}_{\mathcal{D}}(\boldsymbol{\theta})]_{11}},
\end{equation}
where $[\mathcal{F}_{\mathcal{D}}(\boldsymbol{\theta})]_{11}= [J_{\mathcal{D}}^T \mathcal{F}(\boldsymbol{\theta'}) J_{\mathcal{D}} ]_{11}$ is the first (upper left) term of the FIM $\mathcal{F}_{\mathcal{D}}(\boldsymbol{\theta})$ given in Appendix \ref{ax:FIM} when only diffused light is considered, whereas $[\mathcal{F}_{\RegBD}(\boldsymbol{\theta})]_{11} = [J_{\RegBD}^T \mathcal{F}(\boldsymbol{\theta'}) J_{\RegBD} ]_{11}$ stands for the first term of $[\mathcal{F}_{\RegBD}(\boldsymbol{\theta})]_{11}$ obtained when ballistic and diffused light are simultaneously taken into account.  In the above expressions, $J_{\RegBD}$ is the Jacobian of the transformation $\boldsymbol{\theta'} \to \boldsymbol{\theta}$ in the presence of ballistic and diffuse light. We do not require here to compute neither the entire Jacobian $J_{\RegBD}$ or the entire FIM $\mathcal{F}_{\RegBD}(\boldsymbol{\theta})$ as our problem is restricted to finding the first term in the FIM, which provides a limiting but reasonable condition for achieving ballistic filtering when other parameters like $R_\delta$ and $R_{*}$ are already known or assumed to be known. Consequently, the gain can be calculated as
\begin{equation} \mathcal{G}_{bf} =\frac{\frac{1}{\Noise_{\RegBD}} \left(\PartialD{A_{\RegBD}}{M}\right)^2 }{ \frac{1}{\Noise_{\mathcal{D}}} \left(\PartialD{A_{\mathcal{D}}}{M} \right)^2},
\end{equation}
where the $A_{\RegBD}$ and $A_{\mathcal{D}}$ are the amplitudes received at the regions $\RegBD$ and $\mathcal{D}$, respectively (see Appendix~\ref{ax:Amplitudes}). Finally, the expression for the gain $\mathcal{G}_{bf}$ can be written as

    \begin{subequations}
    \begin{eqnarray} \label{eq:ballisticGain}
    \mathcal{G}_{bf} &=&\frac{1}{1+\alpha} \left( 1 + \alpha^2 \beta^2 + 2 \alpha \beta \cos{\Delta\phi} \right)\\
    &=& \frac{1+\frac{\Omega'~e^{ \tau} \left(1+6 e^{-\tau/2} \cos{\sqrt{-1+q^2} R_{\delta }} R_{\text{*}}\right)}{9R_{\text{*}}^2}}{1+\alpha},    
    \end{eqnarray}
    \end{subequations}
where $\tau = 2(q R_{\delta }-R_{*})$, $\beta = \ExpS{R_\delta (q-1)}$ and $\alpha=I_B/I_D$ is the ratio of received ballistic light over diffused light.


    \begin{figure}
        \centering    
        \def\svgwidth{0.95\linewidth}
        \fbox{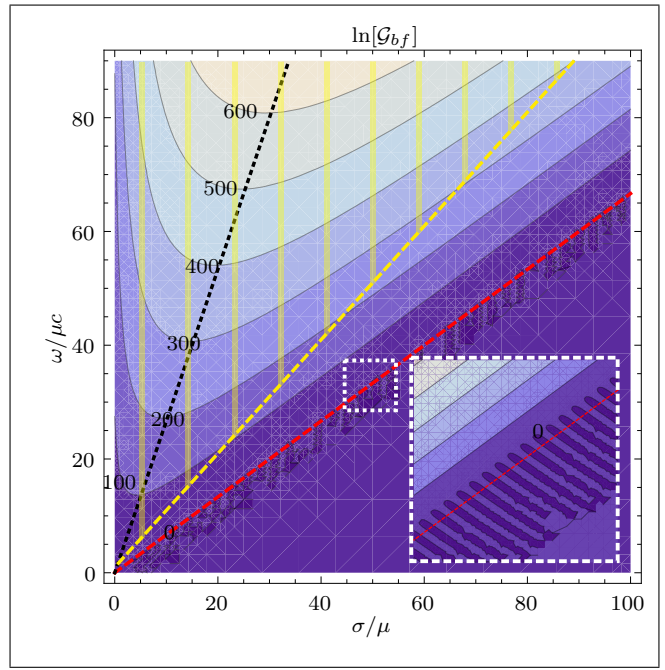}        
        \caption{ Contour plot of $\ln [\mathcal{G}_{bf}]$ for range $\sigma/\mu \in [1, 100]$ and $\omega/ \mu c \in [0.01, 90]$. (Inset) Shows a zoomed in section of the plot where the effect of the cosine term is clearly visible. The cosine term makes it difficult to analytically obtain the contour of unity gain but the condition of Eq.~(\ref{freq_gain_cond}) for expecting a gain is displayed as red dashed line. The diffusion approximation remains valid in the unshaded region below the yellow dashed line.}\label{fig:BallisticFiltering}
        \end{figure}

\subsection{Condition for ballistic filtering gain}

The contour plot of the $\ln [\mathcal{G}_{bf}]$ as a function of the
normalized angular modulation frequency ($\omega/\mu c$) and of
$\sigma/\mu$ is shown in \Fig{fig:BallisticFiltering}, taking
$\Omega' =1$ (isotropic emitter) and $r \mu = 10$. It can be seen that
a gain much greater than unity can be obtained for angular frequencies
that lie above a roughly linear contour. Thus, the minimum angular
frequency required for ballistic filtering depends roughly linearly on
the scattering coefficient of the medium. The contour for unity gain
is not well defined because of the oscillating cosine term in
\Eqn{eq:ballisticGain}, as can be seen in the inset of
\Fig{fig:BallisticFiltering}. Taking a closer look at the expression
of ballistic gain, it can be noticed that a significant gain can be
obtained only when $\tau > 0$, which allows exponential increase in
gain and at the same time exponential decrease in the amplitude of the
cosine term. Moreover, it can be easily checked that it is a
sufficient condition to ensure that $\mathcal{G}_{bf}$ of
Eq.~(\ref{eq:ballisticGain}) is greater than unity. The above simple
condition can be rewritten as $q R_\delta > R_{*}$, which
interestingly suggests that $q R_\delta$ effectively behaves as a
frequency-dependent attenuation for diffused light, which should be
greater than the effective attenuation experienced by ballistic light
($R_{*}$) in order to achieve efficient ballistic filtering.

The above inequality leads to a condition in terms of minimum
frequency of operation which can be simply written as
$q > R_{*}/R_\delta =\sqrt{\bigl(1+\sigma/\mu \bigr)/3}$, which
interestingly appears to be independent of the propagation distance
$r$. The same condition can be expressed in terms of angular frequency
$\omega$, yielding the following gain condition
\begin{equation}\label{freq_gain_cond}
    \frac{\omega}{\mu c} > \frac{2
      \sigma}{3\mu} \sqrt{1 - \frac{\mu}{\sigma} - \frac{2 	\mu^2}{\sigma^2}}.
\end{equation} 
This condition is plotted as red dashed line on the contour map in
\Fig{fig:BallisticFiltering} and seems to provide a well-defined
condition for attaining ballistic gain. Further simplification holds
under the validity conditions of the diffusion theory, where
$\sigma >10 \mu$. In this case, the condition for achieving ballistic
filtering reduces to $ \frac{\omega}{c} > \frac{2}{3} \sigma,$ and is
insensitive to the value of the absorption coefficient $\mu$.

As a result, the above equation provides a simple rule of thumb for
achieving ballistic filtering in the context of intensity modulation
and quadrature detection. It must be noted that the effect of
$\Omega'$ and $r \mu$ (that are set as constants in
\Fig{fig:BallisticFiltering}) is only on the value of the gain and
they do not affect the above condition for efficient ballistic
discrimination.
Identification of this minimum modulation frequency for ballistic filtering is also important from a practical point of view as it is easier to design electronic or electro-optic devices that work at low modulation and demodulation frequencies. The expression derived above can also serve inversely to provide the range of visibilities that can be handled by a ballistic filtering device working at any fixed modulation-demodulation frequency. 

\subsection{Maximum expectable gain}

Lastly, we can estimate the maximum expectable gain under the
condition of highly diffusing medium with reduced ballistic
contribution ($\alpha \ll 1$). The gain values derived below may not
be quantitatively relevant for practical experiments because many
phenomena have been neglected in our analysis so far, such as detector
noise, turbulence, spurious ambient illumination, limited dynamics of
the detector, etc. Under the above conditions, the maximal expected
gain using an intensity modulation scheme and a quadrature
demodulation technique is roughly driven by the exponential term,
i.e., $\ln [{\cal G}_{bf}] \sim \tau$, which allows one to fairly
retrieve the gain values plotted in
Fig. \ref{fig:BallisticFiltering}. As noted above, the value of the
maximum expectable gain depends on $\omega$ and $\sigma$ but also on
$r$ and $\mu$. 
For example, with $\mu = 0.002 ~\text{m}^{-1}$, $\sigma/\mu = 15$ and $r = 1$ km, the minimum angular frequency is $\omega/\mu c = 10$. When operating at 1.5 times this frequency one obtains a gain exponent of 5.    

The evolution of the maximum gain with the physical parameters
involved can be analyzed from the expression of $\tau$. It is obvious
to see that $\mathcal{G}_{bf}$ naturally increases with $\omega$, but
also with $r$. This can be interpreted by noticing that increasing the
number of spatial periods along the propagation must increase the
efficiency of the ballistic filtering. It is also quite
straightforward to show that $\mathcal{G}_{bf}$ increases with the
absorption coefficient $\mu$ when the diffusion approximation is valid
($\sigma/\mu >
10$). 
Though difficult to interpret, it is also possible to demonstrate from
the expression of $\tau$ that $\ln [\mathcal{G}_{{bf}_{max}}]$
increases with $\sigma$ when $\omega / \mu c > 8\sigma/3\mu$ (which
condition is plotted in Fig.~\ref{fig:BallisticFiltering} in dotted
black line), otherwise it should decrease with $\sigma$. However, it
is important to note here that the diffusion equation for modulated
photon transport is no more valid for frequencies higher than
$\omega/c > \mu + \sigma$ \cite{GigahertzFishkin1996}. As a result,
the minimum frequency condition for achieving ballistic filtering (red
dashed line in \Fig{fig:BallisticFiltering}) remains within the domain
of validity, but not the inflection denoted by the black dashed line
in \Fig{fig:BallisticFiltering}.

        Once again, we would like to stress that the quantitative gain
        values retrieved from Fig.~\ref{fig:BallisticFiltering} are
        highly unlikely to be achieved in a practical experiment
        because we have neglected all sources of experimental
        imperfections in our work, and, for very high frequencies, the
        extrapolated gains are obtaind from the diffusion
        approximation beyond its validity. However, the above study
        makes it possible to understand the interplay of the main
        physical parameters at stake in ballistic filtering for
        contrast enhancement.

\section{Conclusion}

We have used the photon diffusion theory and its predictions for transport of intensity modulated light in a diffusive medium along with the noise model for quadrature demodulation scheme, to derive optimal operating points for two application scenarios: scattering parameter estimation as used in diffuse optical imaging and ballistic photon filtering as used in ballistic photon imaging. In the case of estimation of scattering parameters using only diffused light, the Cramer-Rao lower bound on the estimation of scattering parameters was derived and was shown to have a minimum at an optimal modulation frequency. The derived optimal frequency of modulation that achieves minimum variance of estimation depends on the optical penetration depth and the distance of propagation. The evolution of this optimal operating frequency was analytically shown to be increasing with higher absorption coefficient and decreasing with increase in distance of detection and/or the scattering coefficient of the medium. The loss in estimation precision incurred when using a non-optimal operating frequency was quantified. The derivation of the optimal operating frequency paves way for better design of diffuse optical imaging setups that are used in medical instruments.

In the case of ballistic photon imaging using intensity modulated light, an information theoretical metric was introduced to quantify the efficiency of discriminating ballistic photons from diffused photons. A minimum operating angular frequency was derived which appeared to be essentially a linear function of the scattering coefficient of the intervening medium only: it was shown that a significant gain in ballistic filtering can be expected when the angular frequency of modulation $\omega > 2\sigma c /3$. The theoretical expression of the expected gain at any frequency was also derived and its evolution with respect to physical parameters at hand was discussed.

In this approach, diffusion theory of photon transport and noise model
of a quadrature demodulation scheme were tied together using
information theoretical tools to provide minimum\-variance operating
points and to derive the expression of expected gains by taking into
account various competing phenomena in the system. It must however be
kept in mind that the analysis is valid within the applicability of
the diffusion approximation, which down at large angular modulation
frequencies close to $\omega/c = \mu + \sigma$. Moreover, the
extremely high numerical values of gain presented in the article are
only limited by physical photon noise that is carried forward to the
detection scheme. The properties of the detector, like detection noise
and detector dynamics were ignored so far to limit the calculations
and to focus on the physical limits and optimal operation of such a
scheme. However, in real life application, additive detector noise and
source phase noise must be taken into account. Incorporating such
additional noise factors in our approach is a clear perspective to
this work, which will indeed limit the attainable gain values to more
realistic values.

\section*{Acknowledgment}
The authors would like to thank Prof. Afshin Daryoush for fruitful discussions.  

\section*{Funding Information}
This work is funded by CEFIPRA under the project RITFOLD 4604-4 as a collaboration between Institut de Physique de Rennes, Universit\'e de Rennes 1, France and Raman Research Institute, Bangalore, India.  

\noindent


\appendix       
\section*{Appendix}

\section{Noise model}


\subsection{Joint PDF of amplitude and phase} \label{ax:JointPDF}
Let us consider quadrature components $[U,V]^T$ as joint Gaussian random variables with mean $[A \Cos{\phi},A \Sin{\phi}]^T$ and covariance matrix $\Sigma=\mathrm{Diag.}  \left([\Noise , \Noise ]\right)$. Then, the joint distribution of the random variables $U$ and $V$ is
\begin{equation}
\begin{aligned} 
  &P_{U,V}(u,v|A,\phi,\Noise) = \\
  &\frac{1}{2\pi\Noise}\Exp{- \frac{(u - A \Cos{\phi})^2 + (v - A \Sin{\phi})^2}{2\Noise}} .
\end{aligned}
\end{equation} 

For a change of variables, such that $Z = \sqrt{U^2 + V^2}$ and $\Psi = \text{tan}^{-1}[V/U] \implies U = Z \Cos{\Psi}, V = Z \Sin{\Psi} $, changed PDF can be obtained by noting that
\begin{equation}
\begin{aligned}
&P_{Z,\Psi}(z,\psi| A,\phi,\Noise) = P_{U,V}(u,v| A,\phi,\Noise) \left| J^{z,\psi}_{u,v}\right| \\
&= \frac{z}{2 \pi \Noise} \Exp{-\frac{1}{2\Noise} (z^2 + A^2 - 2 z A \Cos{\psi - \phi})},
\end{aligned}
\end{equation}
where, $J^{z,\psi}_{u,v}$ is the Jacobian of the transformation $\{u,v\} \to \{z,\psi\}$ and $|.|$ is the determinant.

\subsection{Noise variance} \label{ax:Noise_variance} At each time slice $t_i$ the photons received at the detector can be modeled as having Poisson noise with variance equal to the mean. Then, the optical noise variance at each slice is $I(t_i) = I_0(1 + M \Cos{2\pi t_i /T})$. The quadrature components are obtained by weighing each slice with sine and cosine of same frequency. Thus, the total noise variance will propagate as \begin{equation*}
\begin{aligned}
&\mathrm{var}(V) = \sum_{i=0}^{nT} \Sin{2\pi t_i/T + \phi_r}^2 I(t)\\
&= \frac{I_0}{2} \\&+ I_0 M \sum_{i=0}^{nT}  (\Sin{2\pi t_i/T + \phi_r} (\Sin{4\pi t_i/T + \phi_r} + \Sin{\phi_r}) )\\
&= \frac{I_0}{2}&
\end{aligned}
\end{equation*}
Similarly, one shows $\mathrm{var}(U) = \frac{I_0}{2}$.
\medskip

\section{Jacobian matrix of the transformation $\boldsymbol{\theta'}
  \to \boldsymbol{\theta}$ for diffuse light only }\label{ax:Jacobian}
The Jacobian matrix for $\boldsymbol{\theta'} \to \boldsymbol{\theta}$ 
\begin{equation} J_{\cal D} =\left[
\begin{array}{ccc}
 \frac{3}{2} e^{-q R_{\delta }} R_{\text{*}} S_0 & -\frac{3}{2} e^{-q R_{\delta }} M q R_{\text{*}} S_0 & \frac{3}{2} e^{-q R_{\delta }} M S_0 \\
 0 & \sqrt{-1+q^2} & 0 \\
 0 & -\frac{3}{2} e^{-R_{\delta }} R_{\text{*}} S_0 & \frac{3}{2} e^{-R_{\delta }} S_0 \\
\end{array}
\right].
\end{equation}

\section{Fisher information $\mathcal{F}_\mathcal{D}(\boldsymbol{\theta})$}\label{ax:FIM}
In the presence of diffused light only, the FIM for a change of coordinates $\boldsymbol{\theta'} \to \boldsymbol{\theta}$ is simply calculated by $\mathcal{F}_\mathcal{D}(\boldsymbol{\theta}) = J_{\cal D}^T~\mathcal{F}(\boldsymbol{\theta'})~J_{\cal D}$
\vfill

 \begin{widetext}
 \begin{equation}
  \mathcal{F}(\boldsymbol{\theta}) = 
  \left[
\begin{array}{ccc}
 \frac{3}{2} e^{(1-2 q) R_{\delta }} R_{\text{*}} S_0 & -\frac{3}{2} e^{(1-2 q) R_{\delta }} M q R_{\text{*}} S_0 & \frac{3}{2} e^{(1-2 q) R_{\delta }} M S_0 \\
 -\frac{3}{2} e^{(1-2 q) R_{\delta }} M q R_{\text{*}} S_0 & 1+\frac{3}{2} e^{(1-2 q) R_{\delta }} M^2 \left(-1+2 q^2\right) R_{\text{*}} S_0 & -\frac{1}{R_{\text{*}}}-\frac{3}{2} e^{(1-2 q) R_{\delta }} M^2 q S_0 \\
 \frac{3}{2} e^{(1-2 q) R_{\delta }} M S_0 & -\frac{1}{R_{\text{*}}}-\frac{3}{2} e^{(1-2 q) R_{\delta }} M^2 q S_0 & \frac{2+3 e^{(1-2 q) R_{\delta }} M^2 R_{\text{*}} S_0}{2 R_{\text{*}}^2} \\
\end{array}
\right].
\end{equation}
\end{widetext}

\section{Amplitude detected at ballistic and diffused regions} \label{ax:Amplitudes}
\paragraph{Diffused region $\mathcal{D}$}
When the detector (or a pixel) receives only diffused light, the
quadrature components detected are given by
    \begin{eqnarray*}
    \begin{aligned}
    u_D &= \int I_D (1+m_D \Cos{\omega t - \phi_D}) \Cos{\omega t + \delta\phi} dt \\ &= \frac{I_D m_D}{2} \Cos{\phi_D + \delta\phi} 
    \end{aligned}
\end{eqnarray*}
\begin{eqnarray*}
    \begin{aligned}
    v_D &= \int I_D (1+m_D \Cos{\omega t - \phi_D}) \Sin{\omega t + \delta\phi} dt \\ &= \frac{I_D m_D}{2} \Sin{\phi_D + \delta\phi} .
    \end{aligned}
    \end{eqnarray*}

\paragraph{Ballistic + Diffused region $\RegBD$}
When the detector (or a pixel) receives both contributions of diffused
light and ballistic light, the quadrature components detected have the
following expressions
    \begin{eqnarray*}    
    \begin{aligned}
    u_{\RegBD} = &\int [I_D (1+m_D \Cos{\omega t - \phi_D}) \\ &+ I_B (1+m_B \Cos{\omega t - \phi_B})]  \times \Cos{\omega t + \delta\phi} dt \\
    & = \frac{I_D m_D}{2} \Cos{\phi_D + \delta\phi} + \frac{I_B m_B}{2} \Cos{\delta\phi}
    \end{aligned}\end{eqnarray*}
Similarly,    
\begin{eqnarray*}
    \begin{aligned}
    v_{\RegBD} = \frac{I_D m_D}{2} \Sin{\phi_D + \delta\phi} + \frac{I_B m_B}{2} \Sin{\delta\phi} .
    \end{aligned}
    \end{eqnarray*}

    \paragraph{Amplitudes}
    As a consequence of the above derivations, the amplitudes
    estimated on detectors that receive only diffused light and
    detectors that receive both ballistic and diffused light
    respectively read
    \begin{eqnarray*}
    	\begin{aligned}
      A_{\mathcal{D}}^2 &= \frac{I_D^2 m_D^2}{4}
      \end{aligned}\end{eqnarray*}
    \begin{eqnarray*}
      \begin{aligned}
      A_{\RegBD}^2 =&  \frac{I_D^2 m_D^2}{4} + \frac{I_B^2 m_B^2}{4} +  \frac{I_D I_B m_D m_B}{2} \Cos{\phi_D}\\
      =& \frac{I_D^2 m_D^2}{4} \left( 1 + \alpha^2 \beta^2 + 2 \alpha \beta \Cos{\phi_D} \right).
      \end{aligned}      
    \end{eqnarray*}


\end{document}

%% file: drawing.pdf_tex
\input{plotStyle.pdf_tex}
\begingroup%
  \makeatletter%
  \providecommand\color[2][]{%
    \errmessage{(Inkscape) Color is used for the text in Inkscape, but the package 'color.sty' is not loaded}%
    \renewcommand\color[2][]{}%
  }%
  \providecommand\transparent[1]{%
    \errmessage{(Inkscape) Transparency is used (non-zero) for the text in Inkscape, but the package 'transparent.sty' is not loaded}%
    \renewcommand\transparent[1]{}%
  }%
  \providecommand\rotatebox[2]{#2}%
  \ifx\svgwidth\undefined%
    \setlength{\unitlength}{540bp}%
    \ifx\svgscale\undefined%
      \relax%
    \else%
      \setlength{\unitlength}{\unitlength * \real{\svgscale}}%
    \fi%
  \else%
    \setlength{\unitlength}{\svgwidth}%
  \fi%
  \global\let\svgwidth\undefined%
  \global\let\svgscale\undefined%
  \makeatother%
  \begin{picture}(1,0.53333333)%
    \put(0,0){\includegraphics[width=\unitlength,page=1]{drawing.pdf}}%
    \put(0.28554393,0.39001841){\color[rgb]{0,0,0}\makebox(0,0)[lb]{\smash{}}}%
    \put(0.49768693,0.12795836){\color[rgb]{0,0,0}\makebox(0,0)[lb]{\smash{$r$}}}%
    \put(0,0){\includegraphics[width=\unitlength,page=2]{drawing.pdf}}%
    \put(0.00846561,0.24169312){\color[rgb]{0,0,0}\makebox(0,0)[lb]{\smash{}}}%
    \put(0.01599441,0.25058201){\color[rgb]{0,0,0}\makebox(0,0)[lb]{\smash{$P_0$}}}%
    \put(0.04656085,0.00211639){\color[rgb]{0.50588235,0.50588235,0.50588235}\transparent{0.514}\makebox(0,0)[lb]{\smash{}}}%
    \put(0.18624339,0.44232802){\color[rgb]{0.50588235,0.50588235,0.50588235}\transparent{0.514}\makebox(0,0)[lb]{\smash{}}}%
    \put(0.0175511,0.33184973){\color[rgb]{0,0,0}\makebox(0,0)[lb]{\smash{$\lambda_0 \pm \Delta\lambda$}}}%
    \put(0.4025645,0.28284271){\color[rgb]{0,0,0}\makebox(0,0)[lb]{\smash{$\Omega$}}}%
    \put(0.7916603,0.26189137){\color[rgb]{0,0,0}\makebox(0,0)[lb]{\smash{$d\Omega$}}}%
  \end{picture}%
\endgroup%

%% file: Sin.pdf_tex
\input{plotStyle.pdf_tex}
\begingroup%
  \makeatletter%
  \providecommand\color[2][]{%
    \errmessage{(Inkscape) Color is used for the text in Inkscape, but the package 'color.sty' is not loaded}%
    \renewcommand\color[2][]{}%
  }%
  \providecommand\transparent[1]{%
    \errmessage{(Inkscape) Transparency is used (non-zero) for the text in Inkscape, but the package 'transparent.sty' is not loaded}%
    \renewcommand\transparent[1]{}%
  }%
  \providecommand\rotatebox[2]{#2}%
  \ifx\svgwidth\undefined%
    \setlength{\unitlength}{274.74077148bp}%
    \ifx\svgscale\undefined%
      \relax%
    \else%
      \setlength{\unitlength}{\unitlength * \real{\svgscale}}%
    \fi%
  \else%
    \setlength{\unitlength}{\svgwidth}%
  \fi%
  \global\let\svgwidth\undefined%
  \global\let\svgscale\undefined%
  \makeatother%
  \begin{picture}(1,0.64068349)%
    \put(0,0){\includegraphics[width=\unitlength]{Sin.pdf}}%
    \put(0.18029194,0.56314278){\color[rgb]{0,0,0}\makebox(0,0)[lb]{\smash{$d_1$}}}%
    \put(0.47010769,0.5374346){\color[rgb]{0,0,0}\makebox(0,0)[lb]{\smash{$d_2$}}}%
    \put(0.69267314,0.15482588){\color[rgb]{0,0,0}\makebox(0,0)[lb]{\smash{$d_3$}}}%
    \put(0.84111757,0.20615462){\color[rgb]{0,0,0}\makebox(0,0)[lb]{\smash{$d_4$}}}%
    \put(0.2479751,0.00027639){\makebox(0,0)[lb]{\smash{1}}}%
    \put(0.38571227,0.00027639){\makebox(0,0)[lb]{\smash{2}}}%
    \put(0.5234457,0.00027639){\makebox(0,0)[lb]{\smash{3}}}%
    \put(0.66118278,0.00027639){\makebox(0,0)[lb]{\smash{4}}}%
    \put(0.7989163,0.00027639){\makebox(0,0)[lb]{\smash{5}}}%
    \put(0.93664974,0.00027639){\makebox(0,0)[lb]{\smash{6}}}%
    \put(0.34441188,0.30854819){\color[rgb]{0,0,0}\makebox(0,0)[lb]{\smash{$2A$}}}%
    \put(-0.00385661,0.02026851){\color[rgb]{0,0,0}\makebox(0,0)[lb]{\smash{$0$}}}%
    \put(-0.00385661,0.34928226){\color[rgb]{0,0,0}\makebox(0,0)[lb]{\smash{$I_0$}}}%
  \end{picture}%
\endgroup%

%% file: plots3.pdf_tex
\input{plotStyle.pdf_tex}
\begingroup%
  \makeatletter%
  \providecommand\color[2][]{%
    \errmessage{(Inkscape) Color is used for the text in Inkscape, but the package 'color.sty' is not loaded}%
    \renewcommand\color[2][]{}%
  }%
  \providecommand\transparent[1]{%
    \errmessage{(Inkscape) Transparency is used (non-zero) for the text in Inkscape, but the package 'transparent.sty' is not loaded}%
    \renewcommand\transparent[1]{}%
  }%
  \providecommand\rotatebox[2]{#2}%
  \ifx\svgwidth\undefined%
    \setlength{\unitlength}{566.1994081bp}%
    \ifx\svgscale\undefined%
      \relax%
    \else%
      \setlength{\unitlength}{\unitlength * \real{\svgscale}}%
    \fi%
  \else%
    \setlength{\unitlength}{\svgwidth}%
  \fi%
  \global\let\svgwidth\undefined%
  \global\let\svgscale\undefined%
  \makeatother%
  \begin{picture}(1,0.48041554)%
    \put(0.81763245,0.45553864){\makebox(0,0)[lb]{\smash{}}}%
    \put(0.86266959,0.45553864){\makebox(0,0)[lb]{\smash{}}}%
    \put(0.01952746,0.51822444){\color[rgb]{0,0,0}\makebox(0,0)[lb]{\smash{}}}%
    \put(0,0){\includegraphics[width=\unitlength,page=1]{plots3.pdf}}%
    \put(0.0151263,0.33331203){\color[rgb]{0,0,0}\rotatebox{90}{\makebox(0,0)[lb]{\smash{$\frac{\omega_{opt}}{\mu c}$}}}}%
    \put(0,0){\includegraphics[width=\unitlength,page=2]{plots3.pdf}}%
    \put(0.03061868,0.27362846){\color[rgb]{0,0,0}\makebox(0,0)[lb]{\smash{1.0}}}%
    \put(0,0){\includegraphics[width=\unitlength,page=3]{plots3.pdf}}%
    \put(0.03061868,0.32450057){\color[rgb]{0,0,0}\makebox(0,0)[lb]{\smash{2.0}}}%
    \put(0,0){\includegraphics[width=\unitlength,page=4]{plots3.pdf}}%
    \put(0.03061868,0.37537128){\color[rgb]{0,0,0}\makebox(0,0)[lb]{\smash{3.0}}}%
    \put(0,0){\includegraphics[width=\unitlength,page=5]{plots3.pdf}}%
    \put(0.03061868,0.42624337){\color[rgb]{0,0,0}\makebox(0,0)[lb]{\smash{4.0}}}%
    \put(0,0){\includegraphics[width=\unitlength,page=6]{plots3.pdf}}%
    \put(0.06621496,0.02337154){\color[rgb]{0,0,0}\makebox(0,0)[lb]{\smash{0}}}%
    \put(0,0){\includegraphics[width=\unitlength,page=7]{plots3.pdf}}%
    \put(0.16487988,0.02337154){\color[rgb]{0,0,0}\makebox(0,0)[lb]{\smash{5}}}%
    \put(0,0){\includegraphics[width=\unitlength,page=8]{plots3.pdf}}%
    \put(0.25830097,0.02337154){\color[rgb]{0,0,0}\makebox(0,0)[lb]{\smash{10}}}%
    \put(0,0){\includegraphics[width=\unitlength,page=9]{plots3.pdf}}%
    \put(0.35696586,0.02337154){\color[rgb]{0,0,0}\makebox(0,0)[lb]{\smash{15}}}%
    \put(0,0){\includegraphics[width=\unitlength,page=10]{plots3.pdf}}%
    \put(0.45628601,0.02337154){\color[rgb]{0,0,0}\makebox(0,0)[lb]{\smash{20}}}%
    \put(0,0){\includegraphics[width=\unitlength,page=11]{plots3.pdf}}%
    \put(0.0336027,0.04550285){\color[rgb]{0,0,0}\makebox(0,0)[lb]{\smash{0.2}}}%
    \put(0,0){\includegraphics[width=\unitlength,page=12]{plots3.pdf}}%
    \put(0.0336027,0.09281695){\color[rgb]{0,0,0}\makebox(0,0)[lb]{\smash{0.4}}}%
    \put(0,0){\includegraphics[width=\unitlength,page=13]{plots3.pdf}}%
    \put(0.0336027,0.14013112){\color[rgb]{0,0,0}\makebox(0,0)[lb]{\smash{0.6}}}%
    \put(0,0){\includegraphics[width=\unitlength,page=14]{plots3.pdf}}%
    \put(0.0336027,0.18744522){\color[rgb]{0,0,0}\makebox(0,0)[lb]{\smash{0.8}}}%
    \put(0,0){\includegraphics[width=\unitlength,page=15]{plots3.pdf}}%
    \put(0.0336027,0.23475938){\color[rgb]{0,0,0}\makebox(0,0)[lb]{\smash{1.0}}}%
    \put(0,0){\includegraphics[width=\unitlength,page=16]{plots3.pdf}}%
    \put(0.25649082,0.00533078){\color[rgb]{0,0,0}\makebox(0,0)[lb]{\smash{$R_\delta$}}}%
    \put(0.28685977,0.41782354){\color[rgb]{0,0,0}\makebox(0,0)[lb]{\smash{(a)}}}%
    \put(0.28888361,0.21970549){\color[rgb]{0,0,0}\makebox(0,0)[lb]{\smash{(b)}}}%
    \put(0.00869835,0.11971636){\color[rgb]{0,0,0}\rotatebox{90}{\makebox(0,0)[lb]{\smash{$\frac{CRB_{22}|\omega_{opt}}{CRB_{22}|\omega_{a}}$}}}}%
    \put(0,0){\includegraphics[width=\unitlength,page=17]{plots3.pdf}}%
    \put(0.8973864,0.16793165){\makebox(0,0)[lb]{\smash{250 MHz}}}%
    \put(0.85094548,0.26389876){\makebox(0,0)[lb]{\smash{600 MHz}}}%
    \put(0.64405941,0.40660464){\makebox(0,0)[lb]{\smash{1300 MHz}}}%
    \put(0,0){\includegraphics[width=\unitlength,page=18]{plots3.pdf}}%
    \put(0.5993524,0.02527509){\makebox(0,0)[lb]{\smash{0}}}%
    \put(0,0){\includegraphics[width=\unitlength,page=19]{plots3.pdf}}%
    \put(0.67160257,0.02527509){\makebox(0,0)[lb]{\smash{10}}}%
    \put(0,0){\includegraphics[width=\unitlength,page=20]{plots3.pdf}}%
    \put(0.74981355,0.02527509){\makebox(0,0)[lb]{\smash{20}}}%
    \put(0,0){\includegraphics[width=\unitlength,page=21]{plots3.pdf}}%
    \put(0.82736221,0.02527509){\makebox(0,0)[lb]{\smash{30}}}%
    \put(0,0){\includegraphics[width=\unitlength,page=22]{plots3.pdf}}%
    \put(0.90490915,0.02527509){\makebox(0,0)[lb]{\smash{40}}}%
    \put(0,0){\includegraphics[width=\unitlength,page=23]{plots3.pdf}}%
    \put(0.98245764,0.02527509){\makebox(0,0)[lb]{\smash{50}}}%
    \put(0,0){\includegraphics[width=\unitlength,page=24]{plots3.pdf}}%
    \put(0.56919345,0.04483706){\makebox(0,0)[lb]{\smash{0.0}}}%
    \put(0,0){\includegraphics[width=\unitlength,page=25]{plots3.pdf}}%
    \put(0.57051807,0.12238568){\makebox(0,0)[lb]{\smash{0.1}}}%
    \put(0,0){\includegraphics[width=\unitlength,page=26]{plots3.pdf}}%
    \put(0.56919345,0.19993437){\makebox(0,0)[lb]{\smash{0.2}}}%
    \put(0,0){\includegraphics[width=\unitlength,page=27]{plots3.pdf}}%
    \put(0.56919345,0.27748301){\makebox(0,0)[lb]{\smash{0.3}}}%
    \put(0,0){\includegraphics[width=\unitlength,page=28]{plots3.pdf}}%
    \put(0.56919345,0.35502987){\makebox(0,0)[lb]{\smash{0.4}}}%
    \put(0,0){\includegraphics[width=\unitlength,page=29]{plots3.pdf}}%
    \put(0.56919345,0.43257854){\makebox(0,0)[lb]{\smash{0.5}}}%
    \put(0,0){\includegraphics[width=\unitlength,page=30]{plots3.pdf}}%
    \put(0.55398203,0.23979226){\color[rgb]{0,0,0}\rotatebox{90}{\makebox(0,0)[lb]{\smash{$\mu$}}}}%
    \put(0.65849744,0.45847322){\color[rgb]{0,0,0}\makebox(0,0)[lb]{\smash{Optimal frequency of modulation}}}%
    \put(0,0){\includegraphics[width=\unitlength,page=31]{plots3.pdf}}%
    \put(0.87228879,0.39979555){\color[rgb]{0,0,0}\makebox(0,0)[lb]{\smash{$r = 3$ cm}}}%
    \put(0.87183725,0.37696084){\color[rgb]{0,0,0}\makebox(0,0)[lb]{\smash{$r = 5$ cm}}}%
    \put(0.78707992,0.0029183){\color[rgb]{0,0,0}\makebox(0,0)[lb]{\smash{$\sigma$}}}%
    \put(0.87307649,0.42170339){\color[rgb]{0,0,0}\makebox(0,0)[lb]{\smash{$n=1$}}}%
    \put(0.52715383,0.42162824){\color[rgb]{0,0,0}\makebox(0,0)[lb]{\smash{(c)}}}%
  \end{picture}%
\endgroup%

%% file: FIM_BF.pdf_tex
\input{plotStyle.pdf_tex}
\begingroup%
  \makeatletter%
  \providecommand\color[2][]{%
    \errmessage{(Inkscape) Color is used for the text in Inkscape, but the package 'color.sty' is not loaded}%
    \renewcommand\color[2][]{}%
  }%
  \providecommand\transparent[1]{%
    \errmessage{(Inkscape) Transparency is used (non-zero) for the text in Inkscape, but the package 'transparent.sty' is not loaded}%
    \renewcommand\transparent[1]{}%
  }%
  \providecommand\rotatebox[2]{#2}%
  \ifx\svgwidth\undefined%
    \setlength{\unitlength}{259.2bp}%
    \ifx\svgscale\undefined%
      \relax%
    \else%
      \setlength{\unitlength}{\unitlength * \real{\svgscale}}%
    \fi%
  \else%
    \setlength{\unitlength}{\svgwidth}%
  \fi%
  \global\let\svgwidth\undefined%
  \global\let\svgscale\undefined%
  \makeatother%
  \begin{picture}(1,1.04160497)%
    \put(0,0){\includegraphics[width=\unitlength]{FIM_BF.pdf}}%
    \put(0.22054398,0.19262349){\makebox(0,0)[lb]{\smash{0}}}%
    \put(0.12157022,0.27353784){\makebox(0,0)[lb]{\smash{100}}}%
    \put(0.19957948,0.37635034){\makebox(0,0)[lb]{\smash{200}}}%
    \put(0.22531636,0.49910497){\makebox(0,0)[lb]{\smash{300}}}%
    \put(0.2549537,0.62353784){\makebox(0,0)[lb]{\smash{400}}}%
    \put(0.28412037,0.74861115){\makebox(0,0)[lb]{\smash{500}}}%
    \put(0.31591435,0.87327935){\makebox(0,0)[lb]{\smash{600}}}%
    \put(0.13177855,0.07874232){\makebox(0,0)[lb]{\smash{0}}}%
    \put(0.28690201,0.07874232){\makebox(0,0)[lb]{\smash{20}}}%
    \put(0.45215664,0.07874232){\makebox(0,0)[lb]{\smash{40}}}%
    \put(0.61741127,0.07874232){\makebox(0,0)[lb]{\smash{60}}}%
    \put(0.78266204,0.07874232){\makebox(0,0)[lb]{\smash{80}}}%
    \put(0.93634259,0.07874232){\makebox(0,0)[lb]{\smash{100}}}%
    \put(0.09320988,0.12589895){\makebox(0,0)[lb]{\smash{0}}}%
    \put(0.07295525,0.30953321){\makebox(0,0)[lb]{\smash{20}}}%
    \put(0.07295525,0.49317133){\makebox(0,0)[lb]{\smash{40}}}%
    \put(0.07295525,0.67680559){\makebox(0,0)[lb]{\smash{60}}}%
    \put(0.07295525,0.86043985){\makebox(0,0)[lb]{\smash{80}}}%
    \put(0.5215938,0.99622135){\color[rgb]{0,0,0}\makebox(0,0)[lb]{\smash{$\text{ln}[\mathcal{G}_{bf}]$}}}%
    \put(0.05677179,0.47674642){\color[rgb]{0,0,0}\rotatebox{90}{\makebox(0,0)[lb]{\smash{$\omega/\mu c$}}}}%
    \put(0.52159374,0.04248994){\color[rgb]{0,0,0}\makebox(0,0)[lb]{\smash{$\sigma/\mu$}}}%
    \put(0.81131339,0.35622131){\makebox(0,0)[lb]{\smash{0}}}%
  \end{picture}%
\endgroup%